\documentclass[a4paper,11pt]{article}
\usepackage{pos}
\usepackage{xfrac}

\usepackage{slashed}

\usepackage{sidecap}
\sidecaptionvpos{figure}{t}

\usepackage{lineno}

\title{Towards dynamical simulations with the anisotropic HISQ action}

\author[a,b]{Alexei Bazavov}
\author*[b]{Yannis Trimis}
\author[c,d]{Johannes H. Weber}

\affiliation[a]{Department of Computational Mathematics,
	Science and Engineering,\\
	Michigan State University, East Lansing, MI 48824, USA}

\affiliation[b]{Department of Physics and Astronomy,\\
	Michigan State University, East Lansing, MI 48824, USA}

\affiliation[c]{Institut f\"ur Physik \& IRIS Adlershof, Humboldt-Universit\"at zu Berlin, \\
Zum Gro\ss en Windkanal 2, D-12489 Berlin, Germany}
\affiliation[d]{Institut f\"ur Kernphysik, Technische Universit\"at Darmstadt, \\
Schlossgartenstra\ss e 2, D-64289 Darmstadt, Germany}

\emailAdd{trimisio@msu.edu}

\abstract{The primary goal of this project is the reconstruction of quarkonium spectral functions from thermal lattice correlators, relevant for the study of Quark-Gluon Plasma in heavy-ion collisions. To this end, we pursue the generation of fully dynamical anisotropic HISQ (aHISQ) ensembles, aiming at a physical strange quark and a heavier-than-physical light quark mass, corresponding to a 300~MeV continuum pion mass. We report on tuning the gauge anisotropy and the lattice spacing of anisotropic pure gauge ensembles with the tree-level Symanzik-improved action using the gradient flow and compare various tuning schemes. We also discuss the simultaneous tuning of the strange quark mass and the quark anisotropy with aHISQ, using spectrum measurements on quenched ensembles. We compare different ways to tune the quark anisotropy and  discuss pion taste splittings for aHISQ at anisotropies up to 8. Finally, we present the expressions for the aHISQ fermion force required for dynamical simulations.}

\FullConference{The 41st International Symposium on Lattice Field Theory (LATTICE2024)\\
 28 July - 3 August 2024\\
Liverpool, UK\\}

\begin{document}
\maketitle
\section{Introduction}
\indent Reconstruction of spectral functions of heavy quarkonia for understanding of their melting pattern in Quark-Gluon Plasma has been studied for over two decades \cite{Asakawa:2000tr}. Our approach to this ill-posed problem, as explained in Ref.~\cite{Bazavov:2023ods}, is based on dynamical simulations with Highly Improved Staggered Quarks \cite{Follana:2006rc} with anisotropy (aHISQ).\\
\indent We present updates on the tuning of gauge and fermion anisotropy as well as the results on pion taste splittings measured on pure gauge ensembles. We also discuss the introduction of anisotropy to the gauge and fermion force, as required for dynamical simulations with the RHMC algorithm \cite{Clark:2003na}.

\section{Anisotropy tuning and pion taste splittings in pure gauge theory}
\begin{figure}[h]
	\begin{center}
		\includegraphics[width=0.6\textwidth]{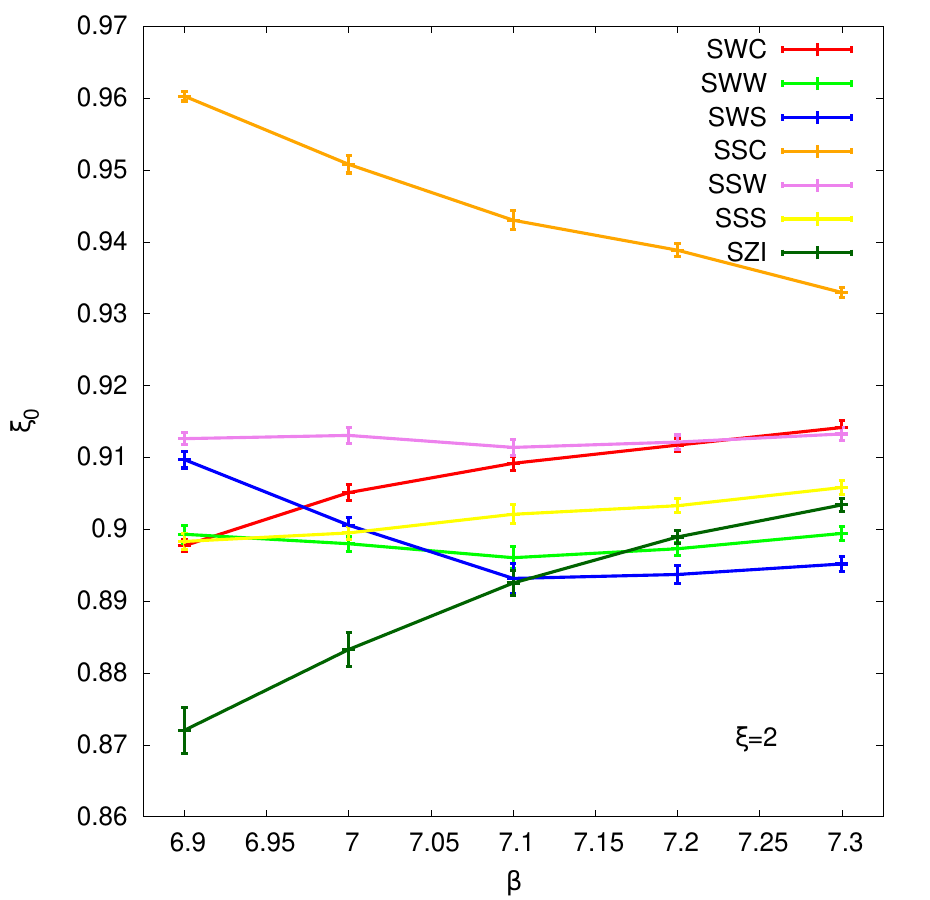}\hfill
		\caption{$\xi_0(\beta)$ at fixed $\xi=2$ for various schemes. ``S'' stands for the tree-level Symanzik-improved action/flow/observable, ``W'' for Wilson, ``Z'' for Zeuthen, ``C'' for clover and ``I'' for improved clover discretization.\label{fig:schemes_xig}}
	\end{center}
\end{figure}

\indent The method we use to tune the lattice spacing and the gauge anisotropy is based on the $w_0$ scale of the gradient flow \cite{Luscher:2010iy} \cite{Borsanyi:2012zr}. Various schemes (combinations of the gauge action, gradient flow action and observable) have different discretization effects as we explored in our previous work~\cite{Bazavov:2023ods}. We continued this exploration by including schemes with Zeuthen flow \cite{Ramos:2015baa} and improved clover operator for the observable, and also by including finer ensembles. 
We tune the lines of constant renormalized anisotropy (LCRA), \textit{i.e.}, find how the bare gauge anisotropy $\xi_0$ changes with $\beta$ such that the renormalized gauge anisotropy is fixed. An example for $\xi=2$ is shown in Fig.~\ref{fig:schemes_xig}.
As one can see in the figure, only the SWC (tree-level Symanzik-improved gauge action, Wilson flow, clover observable) and SZI (tree-level Symanzik-improved gauge action, Zeuthen flow, improved clover observable) schemes are monotonically increasing in the $\beta$ range of interest towards $\xi_0=2$ at $\beta\to\infty$. Non-monotonicity of $\xi_0(\beta)$ may translate to a non-monotonic approach to the continuum limit which we prefer to avoid.

We fit the bare gauge anisotropy as a function of $\beta$ (i.e. $\xi_0=\xi_0(\beta)$ ) at fixed renormalized anisotropy $\xi$ with a Pad\`e functional form:
\begin{equation}\label{eq:la_timide}
	\xi_0(\beta)=\xi\Big(1+\frac{10}{\beta}\ \frac{-a_1+a_2/\beta}{1-a_3/\beta}\Big),
\end{equation}
the same one that was used in Ref.~\cite{Borsanyi:2018srz} (up to rescaling $\beta$ by 5/3 to follow the MILC convention). The fit parameters for $\xi=2$ are
\begin{equation}
	a_1=0.0594\pm0.0004\ ,\quad a_2=0.396\pm0.004\ ,\quad a_3=6.73\pm0.01.
\end{equation}
For comparison, the corresponding fit parameters (in our $\beta$ convention) from Ref.~\cite{Borsanyi:2018srz} are
\begin{equation}
	a_1=0.0578007\ ,\quad a_2=0.375841\ ,\quad a_3=6.5674.
\end{equation}

\begin{figure}[h]
	\begin{center}
		\includegraphics[width=0.6\textwidth]{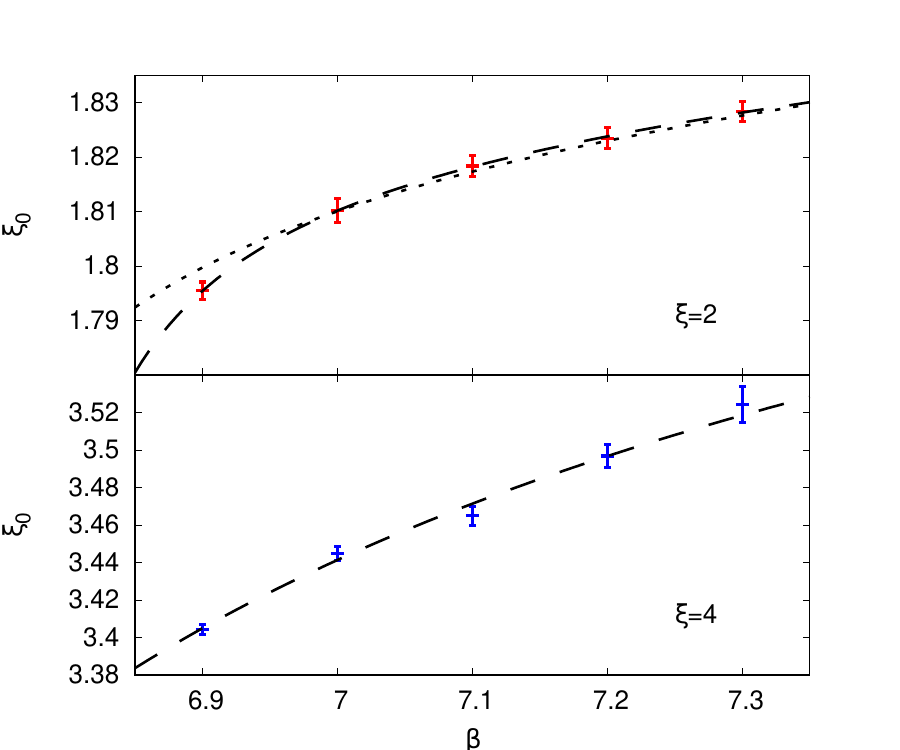}\hfill
		\caption{Pade functional form fitted to data for $\xi=2$ and $\xi=4$ (dashed curves). The dotted curve corresponds to Ref.~\cite{Borsanyi:2018srz}.\label{fig:pade_xig}}
	\end{center}
\end{figure}

Performing the same fit of $\xi_0(\beta)$ for $\xi=4$ LCRA we get
\begin{equation}
	a_1=0.06\pm0.03\ ,\quad a_2=0.3\pm0.3\ ,\quad a_3=6.1\pm0.8.
\end{equation}
The LCRA $\xi=2$ and $\xi=4$ data together with the fits and the fit of Ref.~\cite{Borsanyi:2018srz} for $\xi=2$ are shown in Fig.~\ref{fig:pade_xig}.
One can see that the points for $\xi=2$ follow the curve better than for $\xi=4$, which may be expected since the $\xi=4$ ensembles correspond to coarser lattices at the same values of $\beta$. One can also notice that the fit of Ref.~\cite{Borsanyi:2018srz} (dotted curve) in the upper panel does not pass through the $\beta=6.9$ point, which is reasonable since that work considered only ensembles with $\beta\geqslant7$.

\begin{figure}[h]
	\begin{center}
		\includegraphics[width=0.6\textwidth]{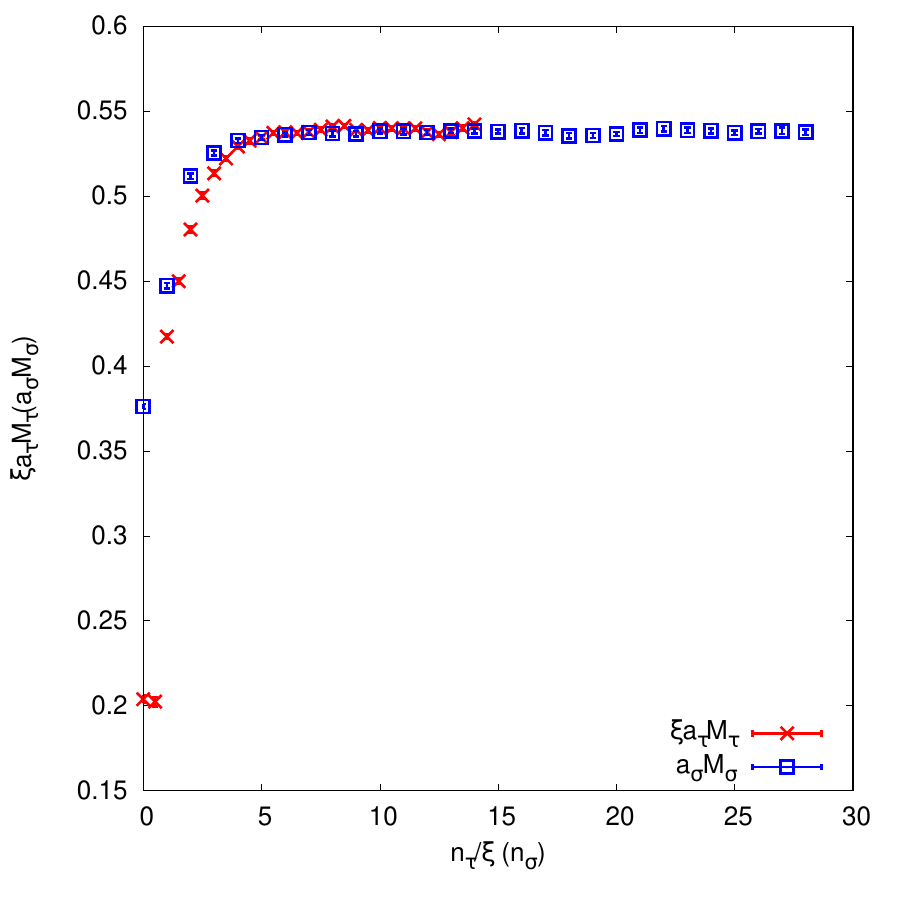}\hfill
		\caption{Effective mass plot for correlators measured in the $\tau$ and $z$ directions on an ensemble with $\xi=2$ and $a=0.16$ fm.\label{fig:rotlat_xig}}
	\end{center}
\end{figure}

\begin{figure}[h]
	\begin{center}
		\includegraphics[width=0.7\textwidth]{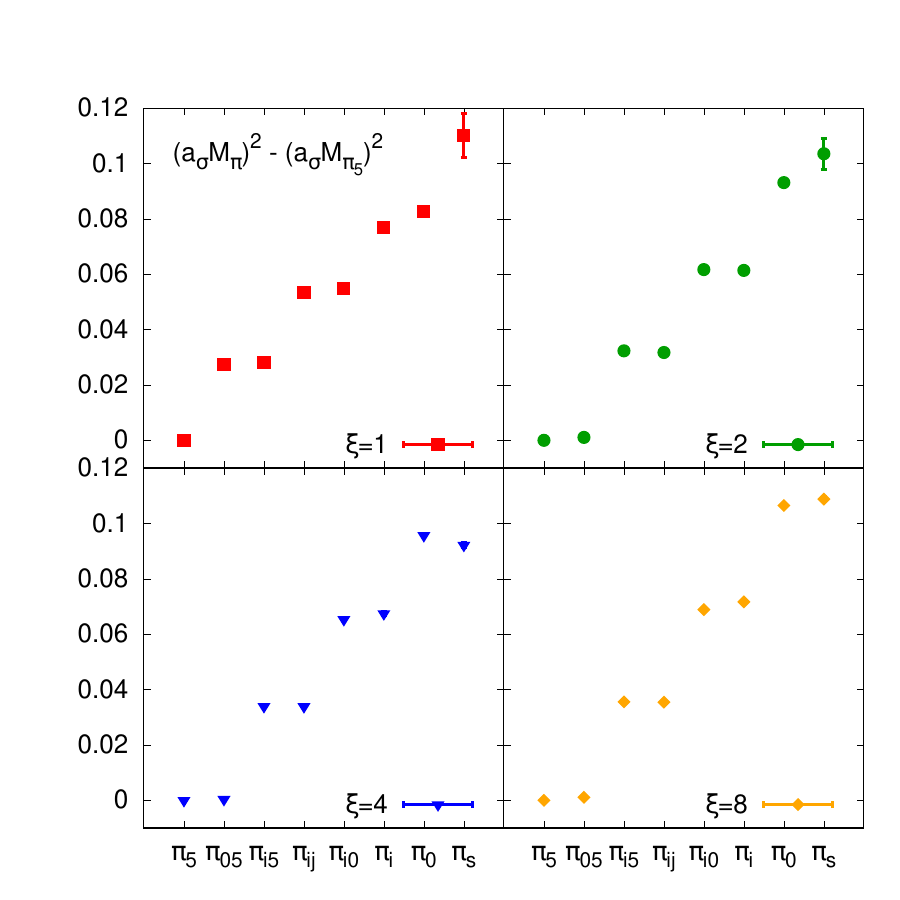}\hfill
		\caption{The pion taste splittings for various anisotropies at $a_\sigma=0.16$ fm.\label{fig:pi_split}}
	\end{center}
\end{figure}

\indent As described in Ref.~\cite{Bazavov:2023ods}, our tuning of the bare fermion anisotropy $\xi_0^f$ is done using the dispersion relation for the fictitious $\eta_{s\bar{s}}$ meson. We experimented, however, with an alternative method where meson correlators are measured in the temporal and in one spatial direction and the ratio of the extracted masses defines the renormalized fermion anisotropy:
\begin{equation}\label{eq:rotlat}
	\xi_f=\frac{a_\sigma M_\sigma}{a_\tau M_\tau}.
\end{equation}
The two effective mass curves are plotted in Fig.~\ref{fig:rotlat_xig}, where horizontal and vertical axes have been rescaled by the renormalized anisotropy $\xi=2$ for the temporal case. The fermion anisotropy for this ensemble was tuned with the dispersion relation method and the coincidence of the two effective mass curves for later times shows the consistence of the two methods. At early times one can notice a minor discrepancy, which stems from the different discretization effects in the two directions.\\ \\
\indent Our results for the pion taste splittings at $a=0.16$ fm for $\xi=1,2,4,8$ are presented in Fig.~\ref{fig:pi_split}. We notice that the known symmetry pattern for staggered mesons at $\xi=1$ \cite{Lee:1999zxa} shifts abruptly to a new pattern at $\xi=2$ that remains as we increase $\xi$.
Our limited studies with naive staggered fermions indicate that in that case the pattern changes more slowly with increasing the renormalized anisotropy $\xi$.

We also note that the $\xi=8$ splittings are preliminary since larger statistics are needed for accurate fits.

\section{Dynamical simulation of aHISQ with RHMC}
\indent Performing dynamical simulations with the anisotropic Highly Improved Staggered Quark (aHISQ) action using the RHMC algorithm \cite{Clark:2003na} within the MILC codebase requires calculation of the force, which determines the updating of the momenta within the molecular dynamics evolution. The force is the sum of the gauge and fermion part.\\
\indent The gauge force is defined as:
\begin{equation}\label{eq:charpentier}
	g_{x,\mu}=\frac{\partial S_g}{\partial U_{x,\mu}}
\end{equation}
where $S_g$ is the anisotropic tree-level Symanzik-improved gauge action:
\begin{equation}\label{eq:gauge_action}
	S_g=\beta\frac{1}{\xi_0}\left[c_0\mathcal{P}_{\sigma\sigma}+c_1\mathcal{R}_{\sigma\sigma}\right]+\beta\xi_0\left[c_0\mathcal{P}_{\sigma\tau}+c_1\mathcal{R}_{\sigma\tau}\right].
\end{equation}
In Eq.~(\ref{eq:gauge_action}) $\mathcal{P}_{\sigma\sigma}$ and $\mathcal{R}_{\sigma\sigma}$ are the sums of purely spatial plaquettes and rectangles, whereas  $\mathcal{P}_{\sigma\tau}$ and $\mathcal{R}_{\sigma\tau}$ the spatial-temporal ones. Coefficients $c_0$ and $c_1$ were determined in Ref.~\cite{Weisz:1982zw}.\\
\indent Substituting Eq.~(\ref{eq:gauge_action}) in (\ref{eq:charpentier}) one can notice that for spatial directions $\mu=1,2,3$ the force gets contributions from both spatial-spatial and spatial-temporal staples, whereas for $\mu=4$
only spatial-temporal staples contribute.\\ \\
\indent Now, for the fermion contribution to the force, we begin by writing down a generic fermion action on an anisotropic lattice:
\begin{equation}
	S_f=a_\sigma^3 a_\tau\sum_x\bar\psi(x)M_{x,y}\psi(y)
\end{equation}
where $M_{x,y}=D_{x,y}+m\delta_{x,y}$. We introduce dimensionless mass $\hat{m}$ as $\hat{m}=a_\sigma m$ and this leads to the dimensionless field definition as $\hat{\psi}=\psi/(a_\sigma\sqrt{a_\tau})$. Then
\begin{equation}
	D_{x,y}=D_{x,y;\sigma}+D_{x,y;\tau}=\frac{1}{a_\sigma}\hat{D}_{x,y;\sigma}+\frac{1}{a_\tau}\hat{D}_{x,y;\tau}=\frac{1}{a_\sigma}\hat{D}_{x,y;\sigma}+\frac{\xi}{a_\sigma}\hat{D}_{x,y;\tau}
\end{equation}
where $\xi=a_\sigma/a_\tau$ is the renormalized anisotropy. Then the action becomes:
\begin{equation}
	S_f=\sum_x\bar{\hat{\psi}}(x)\left[\hat{D}_{x,y;\sigma}+\xi\hat{D}_{x,y;\tau}+\hat{m}\delta_{x,y}\right]\hat{\psi}(y)
\end{equation}
As usual, the parameters $m$ and $\xi$ have to be substituted by their bare values $m_0$ and $\xi^f_0$ in the action, so we finally have:
\begin{equation}
	S_f=\sum_x\bar{\hat{\psi}}(x)\left[\hat{D}_{x,y;\sigma}+\xi_0^f\hat{D}_{x,y;\tau}+\hat{m_0}\delta_{x,y}\right]\hat{\psi}(y).
\end{equation}
It should be mentioned that the bare fermion anisotropy $\xi_0^f$ is different from the bare gauge anisotropy $\xi_0$ as the discretization effects in gauge and fermion sectors are quite different.\\
\indent Following Ref.~\cite{Wong:2007uz} we can now write down the kernel of the anisotropic smeared staggered fermion action:
\begin{align}
	M_{x,y}=2m_0\ \delta_{x,y}+\sum_{\mu=1,2,3}\eta_{x,\mu}\Big[V_{x,\mu}\delta_{x,y-\hat\mu}-V^\dagger_{x-\hat\mu,\mu}\delta_{x,y+\hat\mu}\Big]\nonumber\\
	+\xi_0^f\ \eta_{x,4}\Big[V_{x,4}\delta_{x,y-\hat4}-V^\dagger_{x-\hat4,4}\delta_{x,y+\hat4}\Big]\nonumber\\
	\equiv 2m_0\delta_{x,y}+D_{x,y}.\label{eq:smeared_diracop}
\end{align}
The links $V=V(U)$ are the smeared links that are constructed from the original links $U$; $\eta_{x,\mu}$ are the staggered phases.\\
\indent The fermion force is defined as:
\begin{equation}
	f_{x,\mu}=\frac{\partial S_f}{U_{x,\mu}}.
\end{equation}
Up until Eq.~(2.5) of Ref.\cite{Wong:2007uz} everything is independent of anisotropy. Care is however needed as one makes the step from Eq.~(2.5) to (2.6) of Ref.~\cite{Wong:2007uz} when carrying out the derivatives
\begin{equation}
	\frac{\partial [D^\dagger]_{m,n}}{\partial U_{x,\mu}}\ ,\ \frac{\partial D_{m,n}}{\partial U_{x,\mu}}.\nonumber
\end{equation}
Using the chain rule,
\begin{align}
	\frac{\partial D_{m,n}}{\partial U_{x,\mu}}&=\sum_{y,\nu}\bigg[\frac{\partial D_{m,n}}{\partial V_{y,\nu}}\frac{\partial V_{y,\nu}}{\partial U_{x,\mu}}+\frac{\partial D_{m,n}}{\partial V^\dagger_{y,\nu}}\frac{\partial V^\dagger_{y,\nu}}{\partial U_{x,\mu}}\bigg],\nonumber\\
	\frac{\partial [D^\dagger]_{m,n}}{\partial U_{x,\mu}}&=\sum_{y,\nu}\bigg[\frac{\partial [D^\dagger]_{m,n}}{\partial V_{y,\nu}}\frac{\partial V_{y,\nu}}{\partial U_{x,\mu}}+\frac{\partial [D^\dagger]_{m,n}}{\partial V^\dagger_{y,\nu}}\frac{\partial V^\dagger_{y,\nu}}{\partial U_{x,\mu}}\bigg].\label{eq:forqueray}
\end{align}
The derivatives
\begin{equation}
	\frac{\partial V_{y,\nu}}{\partial U_{x,\mu}}\ ,\ \frac{\partial V^\dagger_{y,\nu}}{\partial U_{x,\mu}}\nonumber
\end{equation}
depend only on the smearing and not on the Dirac operator, so the fermion anisotropy does not affect them. The other four derivatives appearing in Eq.~(\ref{eq:forqueray}) can be calculated using Eq.~(\ref{eq:smeared_diracop}). We find:
\begin{align}
	\frac{\partial D_{m,n}}{\partial V_{y,\nu}}&=\eta_{m,\nu}\delta_{m,y}\delta_{m,n-\hat\nu},\nonumber\\
	\frac{\partial D_{m,n}}{\partial V^\dagger_{y,\nu}}&=-\eta_{m,\nu}\delta_{m-\hat\nu,y}\delta_{m,n+\hat\nu},\nonumber\\
	\frac{\partial D^\dagger_{m,n}}{\partial V_{y,\nu}}&=-\eta_{n,\nu}\delta_{n-\hat\nu,y}\delta_{n,m+\hat\nu},\nonumber\\
	\frac{\partial D^\dagger_{m,n}}{\partial V^\dagger_{y,\nu}}&=\eta_{n,\nu}\delta_{n,y}\delta_{n,m-\hat\nu}\label{eq:rameau_lesplaisirs}
\end{align}
for $\nu\neq4$ and:
\begin{align}
	\frac{\partial D_{m,n}}{\partial V_{y,4}}&=\xi_0^f\ \eta_{m,4}\delta_{m,y}\delta_{m,n-\hat4},\nonumber\\
	\frac{\partial D_{m,n}}{\partial V^\dagger_{y,4}}&=-\xi_0^f\ \eta_{m,4}\delta_{m-\hat4,y}\delta_{m,n+\hat4},\nonumber\\
	\frac{\partial D^\dagger_{m,n}}{\partial V_{y,4}}&=-\xi_0^f\ \eta_{n,4}\delta_{n-\hat4,y}\delta_{n,m+\hat4},\nonumber\\
	\frac{\partial D^\dagger_{m,n}}{\partial V^\dagger_{y,4}}&=\xi_0^f\ \eta_{n,4}\delta_{n,y}\delta_{n,m-\hat4}\label{eq:rameau_lestendresplaintes}
\end{align}
for $\nu=4$.

Substituting Eqs.~(\ref{eq:rameau_lesplaisirs}) and (\ref{eq:rameau_lestendresplaintes}) into Eq.~(\ref{eq:forqueray}) and the result into Eq.~(2.5) of Ref.~\cite{Wong:2007uz} we get an anisotropic generalization of Eq.~(2.6) of Ref.~\cite{Wong:2007uz}
\begin{align}
	[f_{x,\mu}]_{AB}=\sum_{y}(-1)^y\bigg[\sum_{\nu\neq4}\eta_{y,\nu}\Big(\frac{\partial [V_{y,\nu}]_{CD}}{\partial [U_{x,\mu}]_{AB}}[f^{(0)}_{y,\nu}]_{CD}+\frac{\partial [V^{\dagger}_{y,\nu}]_{CD}}{\partial [U_{x,\mu}]_{AB}}[f^{(0)\dagger}_{y,\nu}]_{CD}\Big)\nonumber\\
	+\xi_0^f\ \eta_{y,4}\Big(\frac{\partial [V_{y,4}]_{CD}}{\partial [U_{x,\mu}]_{AB}}[f^{(0)}_{y,4}]_{CD}+\frac{\partial [V^{\dagger}_{y,4}]_{CD}}{\partial [U_{x,\mu}]_{AB}}[f^{(0)\dagger}_{y,4}]_{CD}\Big)\bigg],\label{eq:zelenka}
\end{align}
where
\begin{equation}
	[f^{(0)}_{y,\nu}]_{CD}=\sum_{l}\alpha_l\big([Y^l_{y+\nu}]_D[X^{l*}_y]_C+[X^l_{y+\nu}]_D[Y^{l*}_y]_C\big)
\end{equation}
and
\begin{equation}
	[Y^l_x]_A=[D_{x,y}]_{AB}[X^l_y]_B,\label{eq:lully}
\end{equation}
where now $D$ is the Dirac operator given in Eq.~(\ref{eq:smeared_diracop}). The color indices are explicitly indicated with uppercase roman letters.\\
\indent Factors of $\xi_0^f$ appear explicitly in Eq.~(\ref{eq:zelenka}) but they are also present in $D_{x,y}$ in Eq.~(\ref{eq:lully}) (cf. \ref{eq:smeared_diracop}). Thus, if we rescale all temporal smeared links $V_{x,4}$ by $\xi_0^f$ then Eq.~(\ref{eq:zelenka}) becomes identical to Eq.~(2.6) of Ref.~\cite{Wong:2007uz} and the isotropic RHMC code can be used.

\section{Conclusion}
\indent We explored tuning of the parameters and measured the pion taste splittings for the anisotropic HISQ action in the quenched approximation, reaching renormalized anisotropies up to $\xi=8$. We also introduced anisotropy in the RHMC updating algorithm within the MILC code, which allows us to study its behavior and proceed to the tuning of $2+1$ dynamical aHISQ ensembles, where all parameters-- coupling, quark masses, bare gauge and fermion anisotropies-- have to be tuned simultaneously.



\acknowledgments{
A.B and Y.T.’s research is funded by the U.S. National Science Foundation under the award
No. PHY-2309946. 
J.H.W.’s research has been funded by the Deutsche Forschungsgemeinschaft (DFG, German Research Foundation)---Projektnummer 417533893/GRK2575 ``Rethinking Quantum Field Theory''. 
The authors acknowledge the support by the State of Hesse within the Research Cluster ELEMENTS (Project ID 500/10.006)

The lattice QCD calculations have been performed using the publicly available \href{https://web.physics.utah.edu/~detar/milc/milcv7.html}{MILC code}. Computational resources used in this work were in part provided by the Institute for Cyber-Enabled Research at Michigan State University and the USQCD Collaboration, funded by the Office
of Science of the U.S. Department of Energy.
}

\end{document}